\documentclass[pra,letterpaper,aps,10pt,superscriptaddress,twocolumn,floatfix,showpacs]{revtex4-2}
\usepackage{graphicx}
\usepackage{amsmath}
\usepackage{amsfonts}
\usepackage{float}
\usepackage{amssymb}
\usepackage{epsfig}
\usepackage{epstopdf}
\DeclareGraphicsExtensions{.pdf,.eps,.png,.jpg,.mps}
\usepackage[pdftex]{color}
\usepackage{amsmath,graphicx,amssymb,braket,xcolor,subfigure,upgreek}
\usepackage[colorlinks, linkcolor=blue, citecolor=blue, urlcolor=blue, breaklinks=true]{hyperref}
\usepackage{microtype}
\usepackage{bbm}
\usepackage{color}
\usepackage{dsfont}

\begin{document}

\title{Phase-adaptive cooling of fringe-trapped nanoparticles at room temperature in hollow-core photonic crystal fiber}

\author{Soumya Chakraborty}
\email{soumya.chakraborty@mpl.mpg.de}
\affiliation{Department of Physics, Friedrich-Alexander-Universit\"{a}t Erlangen-N\"urnberg, Staudtstra{\ss}e 7,
	D-91058 Erlangen, Germany}
\affiliation{Max Planck Institute for the Science of Light, Staudtstra{\ss}e 2,
	D-91058 Erlangen, Germany}
\author{Gordon K. L. Wong}
\email{gordon.wong@mpl.mpg.de}
\affiliation{Max Planck Institute for the Science of Light, Staudtstra{\ss}e 2,
D-91058 Erlangen, Germany}
\author{Pardeep Kumar}
\email{pardeep.kumar@mpl.mpg.de}
\affiliation{Max Planck Institute for the Science of Light, Staudtstra{\ss}e 2,
D-91058 Erlangen, Germany}
\author{Hyunjun Nam}
\affiliation{Department of Physics, Friedrich-Alexander-Universit\"{a}t Erlangen-N\"urnberg, Staudtstra{\ss}e 7,
	D-91058 Erlangen, Germany}
\affiliation{Max Planck Institute for the Science of Light, Staudtstra{\ss}e 2,
	D-91058 Erlangen, Germany}
    \author{Claudiu Genes}
\affiliation{TU Darmstadt, Institute for Applied Physics, Hochschulstra{\ss}e 4A, D-64289 Darmstadt, Germany}
\affiliation{Max Planck Institute for the Science of Light, Staudtstra{\ss}e 2,
D-91058 Erlangen, Germany}
\author{Nicolas Y. Joly}
\email{nicolas.joly@mpl.mpg.de}
\affiliation{Department of Physics, Friedrich-Alexander-Universit\"{a}t Erlangen-N\"urnberg, Staudtstra{\ss}e 7,
	D-91058 Erlangen, Germany}
\affiliation{Max Planck Institute for the Science of Light, Staudtstra{\ss}e 2,
D-91058 Erlangen, Germany}

\date{\today}

\begin{abstract}
Active feedback cooling of levitated dielectric particles is a pivotal technique for creating ultrasensitive sensors and probing fundamental physics. Here we demonstrate phase-adaptive feedback cooling of silica nanoparticles optically trapped in standing-wave potential formed by two co-linearly polarized counterpropagating diffraction-free guided modes in a hollow-core photonic crystal fiber at room temperature. Unlike standard laser intensity- or Coulomb force-based feedback, our approach modulates the relative optical phase between the counterpropagating fundamental modes proportionally to the particle's axial momentum. This generates a Stokes-like dissipative force which effectively damps the center-of-mass motion without introducing excess heating and can also work with uncharged particles. At 2 mbar air pressure, the axial center-of-mass temperature of a 195 nm silica particle is reduced by half upon application of the feedback and to 58.6 K at 0.5 mbar. The measured mechanical spectra agree well with our analytical model, validating the cooling mechanism. We envision this approach will open up pathways towards long-range, coherent control of mesoscopic particles inside hollow-core fibers, offering a fiber-integrated versatile platform for future quantum manipulation.
\end{abstract}

\pacs{42.50.Ar, 42.50.Lc, 42.72.-g}

\maketitle

	%
	%
	%
	%
	%
	%
	%
	%
\section{Introduction}
\label{Sec1_Intro}
In recent decades substantial progress has been made on long-standing challenges in both applied and fundamental physics, notably design of sensors with exceptionally high sensitivity and  exploration of quantum mechanics in gravitating objects. Optomechanics \cite{Aspelmeyer_RMP_2014} offers an experimental testbed connecting these seemingly distant branches of physics. As a leading platform in optomechanics, optical levitation in vacuum \cite{Ashkin} allows suspension of solid-state particles of sizes ranging from nanoscale to macroscale using light-induced forces, providing
a high degree of isolation from its thermal environment. These optically trapped mechanical oscillators have become central to a diverse range of applications, including probing limits of quantum mechanics at mesoscopic scales \cite{Millen_ContempPhys_2020}, sensing ultra-weak forces \cite{Ranjit_PRA_2016,Geraci_PRD_2015}, and testing models of wavefunction collapse \cite{Vinante_PRA_2019,Romero_PRA_2011}.

Realizing the full potential of levitated systems requires precise control over the particle’s center-of-mass (CoM) motion, necessitating suppression of both thermal and quantum fluctuations. Various cooling strategies have been developed for this purpose, including passive sideband cooling in high-finesse cavities \cite{Sideband,Uros_Science_2020,Marquardt_PRL_2007,Genes_PRA_2008} and active feedback schemes \cite{Active-feedback} such as parametric feedback \cite{Gieseler_PRL_2012} and cold damping \cite{Tongcang_Li,Sommer_PRL_2019,Sommer_PRR_2020,Tebbenjohanns_PRL_2019}. However, these methods face key limitations. Optical implementations relying on laser \textit{intensity} modulation inherently introduce recoil heating \cite{Recoil_Jain}, while electrostatic cooling techniques \cite{Electric}, though offering an alternative to optical methods, are restricted to charged particles and remain susceptible to stray electric fields. These limitations have motivated the theoretical development of phase-adaptive feedback \cite{Ghosh_PRA_2023, Kumar, Sweden}—a conceptually new cooling technique that dynamically shifts the trapping potential by \textit{phase}-modulating the trapping light in response to the particle’s motion. This technique enables velocity-dependent damping without modifying the trap strength, offering a pathway to  broadband cooling even for uncharged particles. Although its theoretical limits have been established \cite{Kumar}, an experimental realization remains an open problem.

In this paper, we experimentally demonstrate a novel phase-adaptive feedback cooling technique at room temperature, applied to mesoscopic dielectric particles confined by a standing wave trap (SWT) in an antiresonant hollow-core photonic crystal fiber (HC-PCF). The axial fringe-trap is formed due to interference between co-linearly polarized counterpropagating fiber-guided fundamental $\mathrm{LP_{01}}$ modes and the feedback is implemented by dynamically tuning their relative phase based on real-time measurement of the axial momentum of the particle. This phase control induces Stokes-like dissipative forces proportional to the particle’s axial velocity, providing an effective route to suppress thermal noise. Unlike in conventional diffraction-limited optical tweezers, particles in HC-PCF can be trapped over multiple Rayleigh lengths along the fiber \cite{Bykov, Benabid_OptExp_2002, Grass_APL_2016, Linder}, with low leakage loss \cite{lekage_loss_bessel}, and in tunable medium (gaseous, vacuum, or liquid) \cite{Zeltner, Bykov}. Our setup incorporates the distinctive advantages of hollow-core fiber trapping with phase-adaptive feedback control, enabling CoM damping without adding extra heat. Furthermore, the cold particle can be transported through the fiber over ultralong-distance \cite{Linder} along curved trajectories, acting as a long-range point-sensor \cite{Jasper, Bykov} for feeble forces in harsh inaccessible environment. The architecture also supports the trapping of multiple particles in a one-dimensional chain along the fiber axis \cite{Sharma_OptLett_2021}, offering a powerful platform for exploring cooperative phenomena. These results, we believe, will establish a new experimental paradigm for low-noise, long-range control of mesoscopic particles for fiber-based levitated quantum optomechanics.

The paper is organized as follows: Sec.  \ref{Sec2_Exp} describes the experimental setup, including nanoparticle trapping within a HC-PCF and the implementation of phase-adaptive feedback. Cooling results demonstrating axial temperature reduction are presented in Sec. \ref{Sec3_Cooling}, supported by mechanical spectra and a validated stochastic model for the feedback-damped system.  Sec. \ref{Sec4_Discussion} highlights thermal stability advantages over conventional traps and explores future applications in magnetic particle manipulation and force sensing. Finally, concluding remarks are provided in Sec. \ref{Sec5_Conclusion}.

        %
	%
	%
	%
	%
	%
	%
\section{Experiment}
\label{Sec2_Exp}
\begin{figure}[hb!]
\centering 
\includegraphics[width=\linewidth, height=0.9\linewidth]{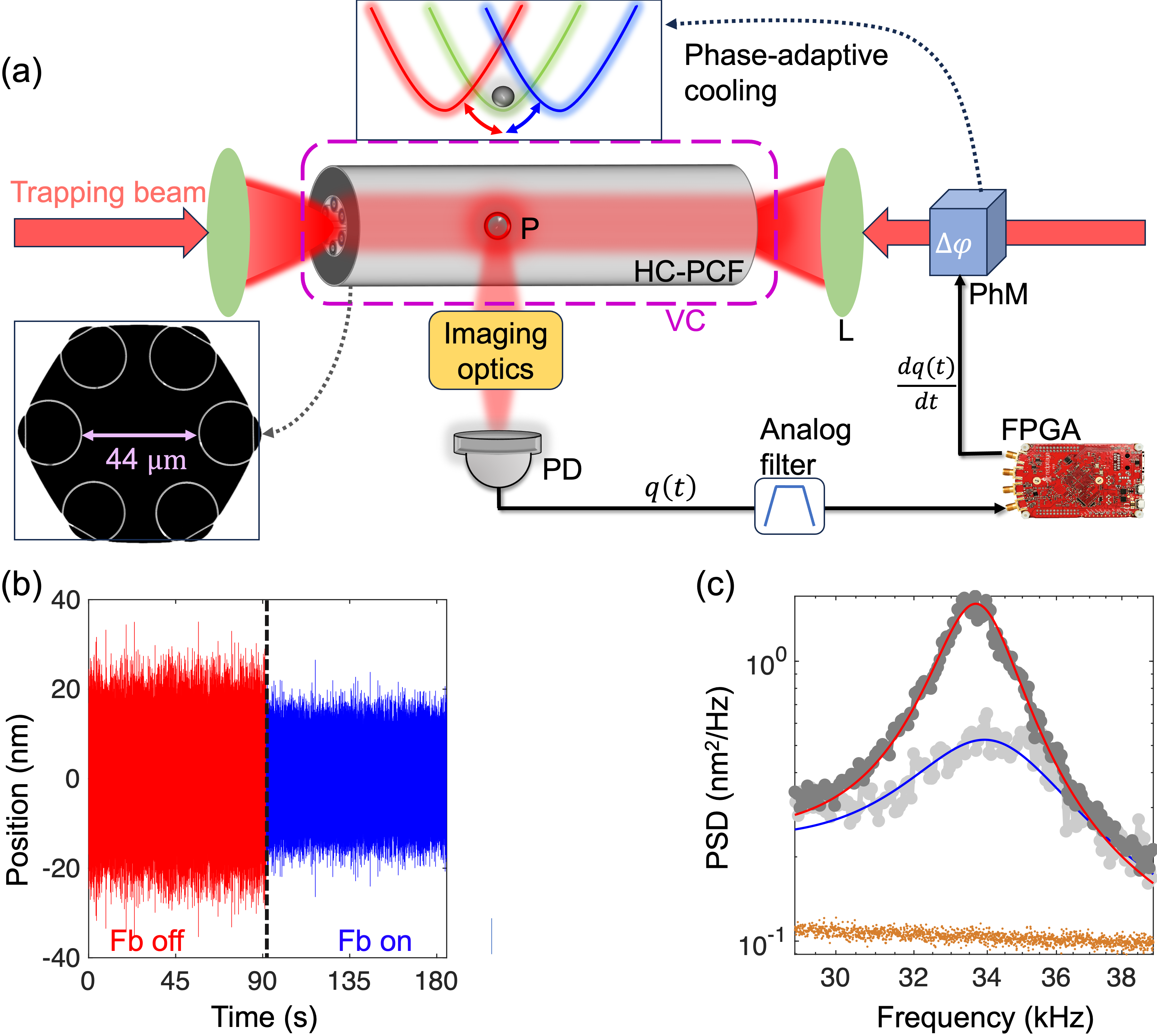}
\caption {A schematic of phase-adaptive active feedback cooling in a dual-beam tweezer setup is shown; P: levitated particle, VC: vacuum chamber, HC-PCF: hollow-core photonic crystal fiber, L: 75 mm lens, PD: photodiode as motion detector, PhM: phase modulator, FPGA: red-pitaya STEMlab-125. The trapping beam is a 1064 nm continuous-wave laser. The total trapping power is 0.75 W. The imaging optics are a combination of spatial aperture, telescope and lenses. The position signal is processed using a 2nd-order Butterworth electronic bandpass filter close to the axial eigenfrequency of the trapped particle. The feedback logic processes time-dependent position ($q(t)$) in order to produce a feedback signal proportional to the filtered axial velocity ($\frac{dq(t)}{dt}$) of the particle. Inset: Scanning electron micrograph of the HC-PCF is shown. Inset: Working principle of phase-adaptive cooling is shown. The optical potential is dynamically adjusted by detuning relative phase between the counterpropagating trapping beams. (b) Long time traces of axial motion of 195 nm silica with (blue) and without (red) feedback (Fb) cooling at 2.0 mbar are shown. The sharp spikes in the time trace are caused by detectors and electronics. (c) Axial mechanical spectra of 195 nm particle at 2.0 mbar with (grey) and without (black) feedback cooling are shown. The axial eigenfrequency is 33.7 kHz. The copper line indicates the noise floor of the detection scheme. The spectra are fitted with Eq.~\eqref{Position_Spectrum}.}
\label{fig1}
\end{figure}

 In Fig.~\ref{fig1}(a), we present the building blocks of the particle trapping and feedback cooling setup. A 10 cm long twisted single-ring HC-PCF \cite{Edavalath:17}, with a core diameter of 44 microns and an outer diameter of 270 microns, is placed inside a vacuum chamber. A scanning electron micrograph (SEM) image of the fiber cross-section is shown in Fig.~\ref{fig1}(a) inset. The twist rate is 0.5 rad/mm and the circular birefringence of the fiber is $10^{-9}$ RIU. The eigenmodes of the twisted fiber are circularly polarized. Thus if co-linearly polarized laser beams are launched from each end of the fiber, the propagating modes will preserve the ellipticity and will suffer identical polarization rotation at every location inside the fiber. This allows generation of standing wave pattern over long-distance along the fiber. Laser light at 1064 nm $(\lambda)$ is coupled to the fundamental $\mathrm{LP_{01}}$ mode of the fiber from each side. The spacing between two adjacent nodes or antinodes is 532 nm. The coherence length of the laser is 5 km ensuring high quality fringe visibility over long distances. The loss of the fiber fundamental mode is 1 dB/m. 

The feedback cooling experiment is performed using silica nanoparticles (Kisker GmbH PSI-5 \% ) of various sizes as long as the particles can be trapped by standing wave fringes (see Appendix \ref{AppendixA}). The initial solution is diluted with isopropanol to a mass concentration of $10^{-8}$ and loaded into the optical trap by aerosol launching technique \cite{Bykov_LSA_2018} until one particle gets trapped in front of the fiber facet. By reducing the power in one of the tweezer arms, the particle is pushed inside the hollow core of the fiber. The scattered light from the particle coming through the semi-transparent fiber cladding is collected using a telescope and a variable spatial aperture. It is simultaneously focused on a fast camera and an InGaAs (Indium Gallium Arsenide) amplified photoreceiver. The aperture size can be adjusted in order to control the amount of scattered light reaching the detector. At low pressure due to stochastic thermal noise, as the particle drifts away from the highest intensity point of the fringe, the brightness of the scattered light changes. The output of the photodiode therefore contains the positional information of the trapped particle. This side-scattered light also contains extraneous stray light that comes out of the fiber cladding, setting up a strong detection noise background.  
The output signal of the photodiode is filtered with a second-order bandpass Butterworth filter close to the axial eigenfrequency and passed through the feedback logic circuit to produce a feedback signal (see Appendix \ref{AppendixB}). In order to control the relative phase between the trapping beams, a lithium niobate phase modulator is installed in one of the tweezer arms. The modulator is driven by the amplified feedback signal using a high-voltage amplifier. 
After trapping a 195 nm silica particle, the fiber is evacuated using a turbo pump. At low pressure inertial motion of the particle becomes underdamped featuring mechanical eigenresonances. The axial resonance frequency ($\Omega$ = 33.7 kHz) is verified by driving the phase modulator with a small-signal sinusoidal perturbation (see Appendix \ref{AppendixC}). We observed slight variation in the trapping frequency (0.8 kHz) over the course of experiment most likely caused by the intensity fluctuations of the trapping laser.
The feedback logic is switched on when the particle's motion becomes ballistic, usually below 5 mbar.

The principle behind the feedback mechanism is illustrated in Fig~\ref{fig1}(a) inset. As the particle drifts away from the potential minima of the axial trap, the fringe position is dynamically adapted by modulating the relative phase between the trapping beams. This, in turn, generates a Stokes-type drag force (see Appendix \ref{AppendixD}), thereby damping the CoM motion. 

        %
	%
	%
	%
	%
	%
	%

\section{Cooling action}
\label{Sec3_Cooling}

The effect of the feedback cooling at 2 mbar along the axial-direction is shown in Fig.~\ref{fig1}(b) and (c). The time-domain traces of oscillation of the particle without (red) and with (blue) feedback cooling are shown in Fig~\ref{fig1}(b). The signal is calibrated using equipartition theorem \cite{Calibration} in the absence of feedback. Upon activation of the phase-adaptive feedback, the RMS amplitude of motion becomes small, indicating an increase in the natural damping rate $\gamma$.

A complementary frequency-domain representation is shown in Fig.~\ref{fig1}(c). The dark-grey curve shows the position power spectral density (PSD) of the particle motion without feedback, while the light-grey curve corresponds to the PSD under feedback cooling. The copper-colored line marks the detection noise floor. The application of feedback results in a broadened spectrum and reduced peak amplitude, reflecting increased damping and suppressed fluctuations. Moreover, the integrated area between the position PSD and the noise floor—proportional to the particle’s potential energy (CoM temperature) along the axial direction—is significantly reduced, confirming that thermal energy is extracted from the system by the feedback mechanism.  In principle, further reduction in temperature is possible until the PSD goes below this noise floor (0.4 nm/$\sqrt{\mathrm{Hz}}$). The detection noise floor can be improved by closing the spatial aperture in front of the photodiode while simultaneously trapping and imaging a bigger particle which scatters more light (see Appendix \ref{AppendixE}).

Our side-scattered detection resembles in-loop detection schemes \cite{Electric}. To accurately model the experimental spectra near resonance, we now introduce the effects of detection and feedback noise in the phase-adaptive cooling protocol. These noise sources originate from the detection and amplification of stray extraneous scattered light by the fiber cladding, as well as intrinsic laser noises. For smaller particles of mass $m$, the scattering cross section decreases as $\sigma_{scat}\propto m^2$, requiring a wider collection aperture to gather sufficient signal for imaging. However, increasing the aperture also admits more stray light, thereby enhancing the detection noise. Ultimately, when feedback is applied, this noise is amplified and re-injected into the trap, manifesting as feedback-induced noise. 

To model the position PSD under phase-adaptive feedback, we describe the dynamics of the trapped particle using coupled stochastic difference equations for the position and momentum quadratures~\cite{Kumar}. These equations are discretized and analyzed in the long-time limit to capture the system's steady-state behavior. In this framework, the detected position signal is expressed as $z_{\mathrm{det}}(t) = z(t) + q_{\mathrm{zpm}} \eta W^{\mathrm{det}}(t)$, where $z(t)$ captures the thermal and feedback-induced fluctuations, while the second term accounts for additive detection noise \cite{Vivishek_Sci_2021} characterized by amplitude $\eta$, Wiener increments $W^{\mathrm{det}}(t)$, and the zero-point motion (zpm) amplitude $q_{\mathrm{zpm}}=\sqrt{\hbar/m\Omega}$. Incorporating these contributions, the PSD near the axial mechanical resonance is derived from the long-time solution of the discretized equations (see Appendix \ref{AppendixD}) and takes the form:
\begin{align}
S(\omega) &= \frac{2\gamma k_B T_{\mathrm{th}}/m}{(\omega^2 - \Omega^2)^2 + (\gamma + \beta \mathcal{C})^2 \omega^2}+\Bigg(\frac{\hbar}{m\Omega}\Bigg)\frac{\eta^{2}}{\omega^{2}}\nonumber
\\&+\Bigg(\frac{\hbar}{m\Omega}\Bigg) \Bigg[\frac{ (\eta \beta \mathcal{C})^2}{(\omega^2 - \Omega^2)^2 + (\gamma + \beta \mathcal{C})^2 \omega^2}\Bigg]\;,
\;\label{Position_Spectrum} 
\end{align}
where $k_B$ is the Boltzmann's constant, $\hbar$ is the reduced Planck's constant and $T_{\mathrm{th}}$ denotes the bath temperature 293 K.  Above spectrum comprises three distinct contributions: the first term represents the desired feedback-damped mechanical response, characterized by the parameter $\beta=\frac{\kappa}{k_{0} p_{\mathrm{zpm}}}=4.49$ $\mathrm{GHz}$, which is associated with the trap stiffness ($\kappa$) and the wavevector ($k_{0}$) of light. Here, $p_{\mathrm{zpm}}=\sqrt{m\hbar\Omega}$ denotes zpm momentum of the axial motion. The parameter $\mathcal{C}$ quantifies the feedback strength, reaching a maximum value \cite{Kumar} of $2\pi\sqrt{\hbar\Omega/k_{B}T_{\mathrm{th}}}$.  The second term accounts for detection noise of magnitude $\eta$, which contributes an apparent amplitude offset near resonance. Its $\omega^{-2}$ scaling arises from additive noise in the position measurement \cite{Vivishek_Sci_2021}, consistent with the discrete-time stochastic model (see SM~\cite{SM}). The third term denotes the undesired feedback noise arising due to the amplified detection noise being fed back into the optical trap. Away from the resonance, the spectrum is limited by the base noise floor of the detector.  

We fit the position PSD in Fig.~\ref{fig1}(c) for feedback-off (solid red line) and feedback-on (solid blue line) conditions using Eq.~\eqref{Position_Spectrum}, extracting $\gamma=2$ kHz, $\eta = 9.4 \times 10^{6}~\sqrt{\mathrm{Hz}}$ and $\mathcal{C} = 3.84 \times 10^{-6}$ through a two-step procedure: First, the feedback-off spectrum determines $\eta$, $\Omega$, and $\gamma$; second, with these fixed, the feedback-on spectrum yields $\mathcal{C}$. The value of $\eta$ is independently verified via thermal occupancy measurements under feedback cooling.

        %
	%
	%

\begin{figure}[htbp]
\centering
\includegraphics [width=\linewidth]{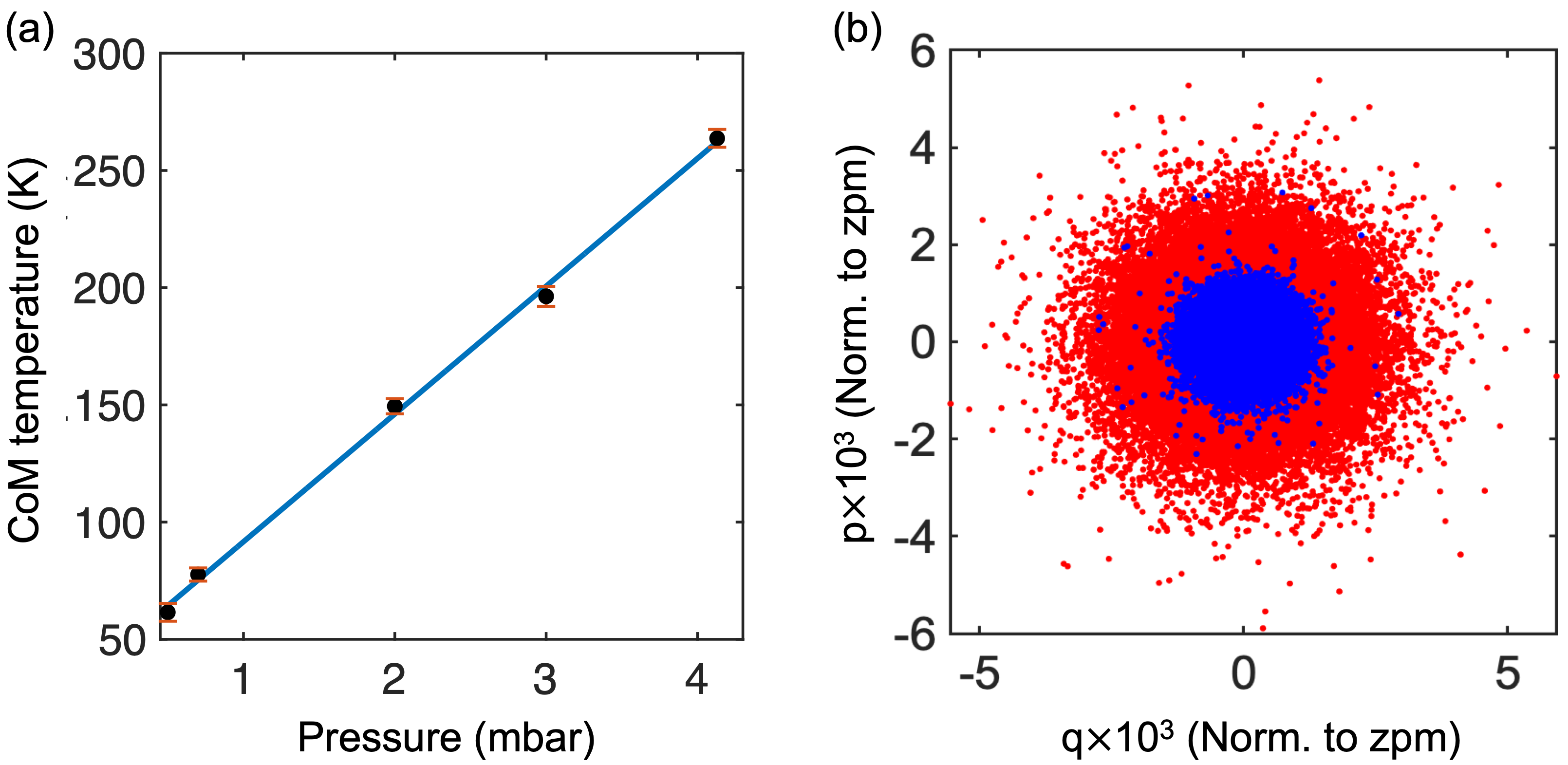}
\caption {(a) CoM temperature as a function of pressure is shown. The CoM temperature is proportional to the integrated area between the measured spectrum and the noise floor, after calibrating them with uncold spectral area at 293 K. The error bar indicates standard deviation of temperature calculated by measuring the spectrum 32 times at a given pressure. (b) Experimental phase-space distributions at 0.7 mbar are shown with (blue) and without (red) feedback cooling. The position ($q$) and momentum ($p$) quadratures are normalized to the zero-point motion (zpm) of the oscillator in motional ground state, where $q_{\mathrm{zpm}}=8.22$ pm and $p_{\mathrm{zpm}}=12.1$ yN/s.}
\label{fig2}
\end{figure}

In our experiment, particle loss typically occurs below 1~mbar due to the absence of full three-dimensional cooling. Nonetheless, in certain cases, stable trapping is maintained down to 0.5~mbar. The CoM temperature of the axial eigenmode is proportional to the integrated area between the measured spectrum and the noise floor, using the feedback-off spectrum as a reference at 293 K. As shown in Fig~\ref{fig2}(a), the temperature decreases linearly with pressure, consistent with the expected suppression of Brownian noise.

The reduction in CoM temperature~\cite{Calibration} can be visualized in phase space (see Fig.~\ref{fig2}(b)). The red and blue distributions correspond to the cases without and with feedback, respectively. These are obtained by sampling the mechanical trajectory over a time interval much longer than the oscillation period, using a sampling rate that exceeds the mechanical decoherence rate, \(1/\gamma\). The experimentally measured phase-space distribution was recorded at a pressure of 0.7\,mbar, based on a continuous time trace exceeding 120\,s. In the absence of feedback (red points), the oscillator exhibits a broad thermal Gaussian distribution, characteristic of the bath temperature \(T_{\mathrm{th}}\). When feedback is applied (blue points), the distribution contracts significantly, indicating a reduction in CoM temperature. This compression in phase space reflects a decrease in thermal energy and remains consistent with energy equipartition, even under externally applied damping.

        %
	%
	%
	%
	%
	%
    \section{Discussion}
\label{Sec4_Discussion}

The ultimate achievable CoM temperature is limited by the loss of the particle in high vacuum, possibly due to an increase in internal bulk temperature of the nanosphere \cite{ricciThesis2019, HebestreitPRA2018, ChangPNAS2010} and insufficient viscous damping at low pressure. Additional damping can be provided in the radial directions by using phase-adaptive parametric feedback \cite{Ghosh_PRA_2023,Gieseler_PRL_2012}, while cooling efficiency can be improved by enhancing the oscillator’s quality factor—for instance, by trapping it in a smaller-core HC-PCF, as the axial eigenfrequency scales linearly with the core diameter \cite{Self}. Furthermore, the detection noise floor can be reduced using homodyne measurement with the particle's self-image \cite{DaniaSelfImage_PRL_2022}.

The demonstrated phase-adaptive cooling scheme, combined with a hollow-core fiber trapping platform, may offer a viable route for cooling magnetic microparticles \cite{Kumar} and for probing weak magnetic forces over extended distances using moving interference fringes as an optical conveyor belt \cite{Cizmar_APL_2005, Self}. The hollow-core fiber enables weak trapping of magnetic particles without the need for tightly focused, diffraction-limited laser beams, which can significantly increase the internal temperature of the particle and, in extreme cases, drive it close to its Curie temperature—thereby compromising its magnetic properties. We calculate that 1 micron yttrium iron garnet (YIG) particle trapped in a diffraction-limited beam using 0.85 NA lens, and 1 W of trapping power can attain 600 K at 0.1 mbar assuming 10 dB/km loss. In contrast, in hollow-core fiber, the corresponding temperature is 340 K at the same pressure level under the same conditions. The Curie temperature of YIG is 560 K. Therefore fringe-trapping in hollow-core fiber offers superior thermal stability of the trapped magnetic particle. 

Heating may also lead to the evaporation of the outer shell of the particle \cite{Millen_NanoTech_2014}, giving rise to large instantaneous momentum kicks and making the trapped particle unstable in high vacuum. The issue of internal temperature can also be tackled by doping the particles \cite{DopeSi} with rare-earth ions, providing pathways to redistribute thermal energy
via spontaneous emission at selected resonances of large optical emission rate (solid-state refrigeration) \cite{RahmanNatPhoton2017}.

        %
	%
	%
	%
	%
	%
\section{Conclusion}
\label{Sec5_Conclusion}
We have experimentally demonstrated a novel phase-adaptive feedback cooling technique, by trapping silica nanoparticles in interference fringes inside the hollow core of a photonic crystal fiber. The cooling method involves modulation of the phase of fringe-trap. This technique may provide a remedy for extraneous heating often associated with traditional techniques involving intensity modulation. This work opens up new opportunities for long-range precision sensing and coherent manipulation of mesoscopic quantum systems.

        %
	%
	%
	%
	%
	%

\section*{Acknowledgments}

We acknowledge financial support from the Max Planck Society and the Deutsche Forschungsgemeinschaft (QuCoLiMa: ID-429529648, OrbitFlySens: ID-418737652). HN acknowledges financial support from Max Planck School of Photonics (MPSP). SC acknowledges many insightful discussions with Carla Maria Brunner throughout the course of this work. SC and GKLW acknowledge Lothar Meier, at MPI for the Science of Light, for his tremendous help and support in fabricating the electronic circuits used in the experiments. SC and HN acknowledges Alekhya Ghosh and Pascal Del'Haye for lending the FPGA. 


        %
	%
	%
	%
	%
	%

 \section*{Data availability} The authors confirm that the theoretical derivations supporting the findings of this study are available within the article and its supplementary materials. The experimental data and numerical analysis is available from the corresponding authors, upon reasonable request.

\appendix

\section{Axial spring constant as a function of particle diameter}
\label{AppendixA}

\begin{figure}[htbp]
\centering
\includegraphics[width=0.6\columnwidth]{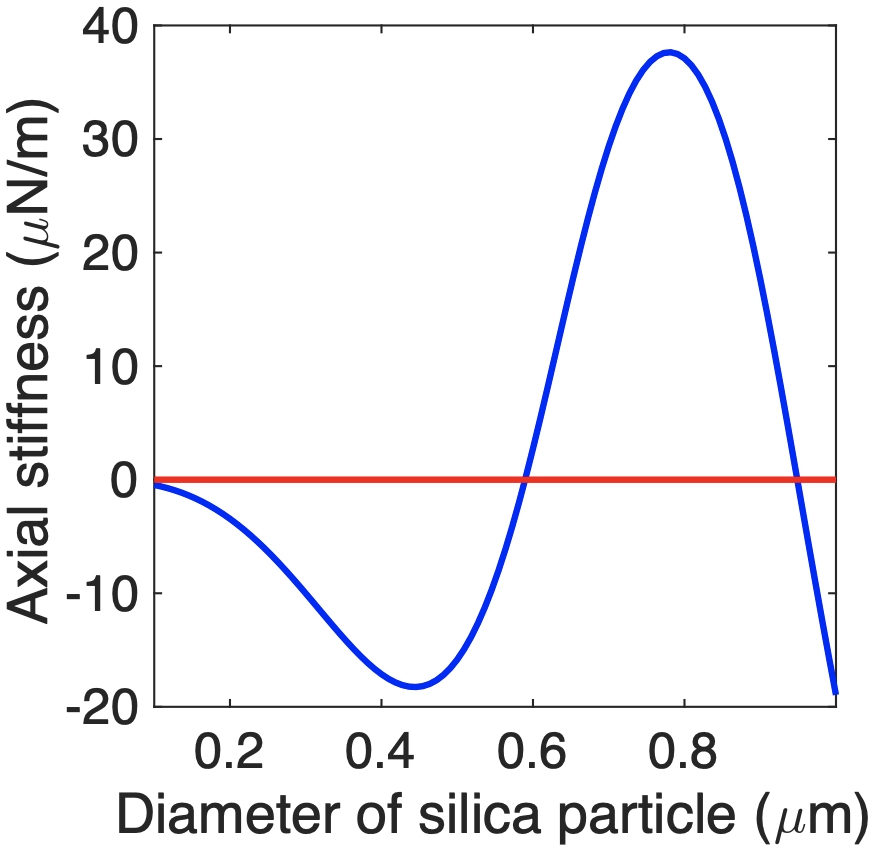}
\caption {Axial stiffness is plotted (blue curve) as a function of silica particle diameter. Parameter used in the simulation are: NA of the fiber = 0.03, density of silica = 2000 $\mathrm{kg}/\mathrm{m}^3$, refractive index of silica = 1.43, total trapping power = 1 W. The simulation is performed using optical tweezer toolbox \cite{Nieminen_2007}. The red line indicates zero stiffness. The crossover between blue and red curve indicates the sizes of the particle which do not feel any axial gradient force and therefore cannot be cooled down.}
\label{stiffness}
\end{figure}
In our experiment, we have trapped and cooled center-of-mass (CoM) motion of silica particles of various sizes. The trapping mechanism of larger particles is governed by generalized Lorentz-Mie theory (GMLT) \cite{GMLT, Nieminen_2007}. The silica particles can be high-field seeker (trapped at antinode) or low-field seeker (trapped at node) depending on their sizes and experimental parameters \cite{Stiffness}. In Fig.~\ref{stiffness}, we show the simulated axial optical spring stiffness (blue curve) as a function of silica particle diameter. It can be seen that the particles larger than fringe spacing can also be cooled down, however the direction of the phase change has to be adjusted depending on the particle diameter. Certain sizes of particles do not feel any axial gradient force (red line), thus cannot be cooled down using phase-adaptive technique.

\section{Methods}
\label{AppendixB}
The feedback cooling logic is implemented by using operational amplifier based on analog feedback in the beginning and later on digital feedback by red pitaya FPGA (STEMlab 125-14). The analog circuit consists of a bandpass filter, a unity gain derivative circuit, a phase-shifter for additional adjustment of the phase delay, an inverter to switch from cooling to heating and a switch for turning off/on the feedback logic. For FPGA implementation, we used IQ-module from Pyrpl package \cite{8087380}. The module performs convolution of the input signal with reference sine and cosine functions and can be modified to be used as a narrow-bandwidth bandpass filter. We set the center frequency ($\Omega$) of the filter to be the axial eigenfrequency of the oscillator. We also provide a suitable phase shift to delay the output by $\frac{\pi}{2\Omega}$ from the input signal so that it can work as a feedback signal in the cold-damping regime. Finally, a gain factor is set to adjust the strength of the applied feedback. Additionally if the oscillator's eigenfrequency is low, one can also use a USB-DAQ card to perform the feedback logic.


\section{Verification of the axial eigenfrequency}
\label{AppendixC}

\begin{figure}[h]
\centering
\includegraphics[width=0.6\columnwidth]{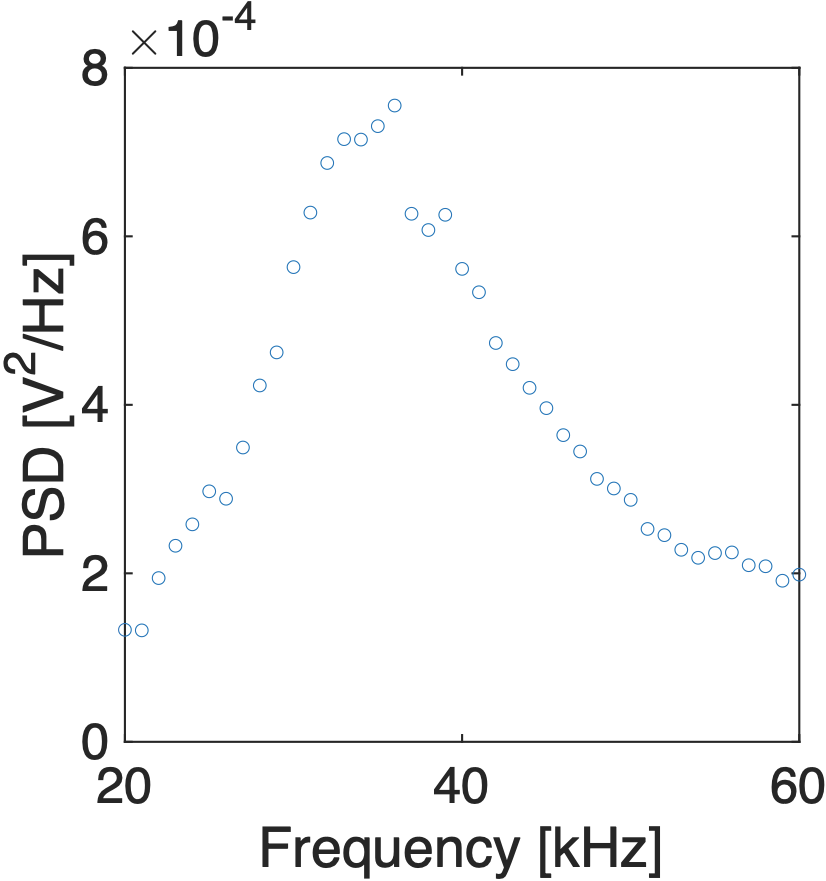}
\caption {Verification of the axial eigenfrequency by driving the standing wave trap with sinusoidal small-signal modulation at 45 mbar is shown. The scattered light power spectral density is recorded with a lock-in detection.}
\label{lockin}
\end{figure}
In order to verify the axial eigenmode, the standing wave trap is sinusoidally perturbed at 45 mbar with a small signal modulation and the scattered light spectrum is recorded using a lock-in detection. The result is shown in Fig.~\ref{lockin} verifying the axial eigenfrequency.    
\section{Derivation of the position spectrum}
\label{AppendixD}
To derive the steady-state position spectrum under phase-adaptive feedback, we follow the formalism introduced in Ref.~\cite{Kumar}. In this framework, the stochastic dynamics of a trapped particle subjected to phase-adaptive feedback along the axial ($z$-) direction, is described by coupled difference equations expressed in terms of dimensionless position and momentum quadratures, defined as \( q = z / q_{\mathrm{zpm}} \) and \( p = p_z / p_{\mathrm{zpm}} \), where \( q_{\mathrm{zpm}} = \sqrt{\hbar / (m \Omega)} \) and \( p_{\mathrm{zpm}} = \sqrt{m \hbar \Omega} \) denote the zero-point position and momentum scales, respectively. The equations of motion read:
\begin{subequations}  
\begin{align}  
dq &= \Omega p  dt, \\  
dp &= -\gamma p  dt - \Omega q  dt - \beta \phi(t)  dt + \sqrt{2\gamma n_{\mathrm{th}}}  dW^{\text{th}}(t)\;.  
\end{align}  
\label{EQsMotion}  
\end{subequations}  
Here, the system is subject to two dominant stochastic processes: (1) Thermal noise from background gas collisions couples the oscillator to a bath at temperature $T_{\text{th}}$ (occupancy $n_{\text{th}}$), introducing pressure-dependent damping ($\gamma$) and fluctuations modeled via Wiener increments $dW^{\text{th}}(t)$ satisfying $\langle (dW^{\text{th}}(t))^2 \rangle = dt$ (time-step); (2) Feedback force $-\beta \phi(t)$ with $\beta = m\Omega^2 \lambda /(2\pi p_{\text{zpm}})$ ($\lambda$: laser wavelength) generated through phase control.

Our phase-adaptive feedback cooling implementation modulates the relative phase between the counter-propagating fiber modes as
\begin{equation}
\phi(t) = \frac{\mathcal{C}}{\Omega} \times \lim_{dt\to 0} \frac{q_{\mathrm{det}}(t) - q_{\mathrm{det}}(t - dt)}{dt}\;,\label{Relative_Phase}
\end{equation}
where the measured position $q_{\mathrm{det}}(t) = q(t) + \eta  W^{\mathrm{det}}(t)$ includes additive detection noise characterized by amplitude $\eta$ and Wiener increments $W^{\mathrm{det}}(t)$. The feedback gain $\mathcal{C}$ governs a critical performance trade-off: increasing $\mathcal{C}$ enhances cooling rates but amplifies noise contributions. Note that, substituting Eq.~\eqref{Relative_Phase} into Eq.~\eqref{EQsMotion} reveals an enhancement of the mechanical damping from $\gamma$ to $\gamma + \beta\mathcal{C}$. This feedback mechanism corresponds to cold damping.

To derive an analytical solution for this system, we discretize the time interval $[0, t]$ into $n$ steps of duration $dt = t/n$, following the effective numerical scheme of Ref.~\cite{Kumar}. This discretization reformulates the stochastic equations of motion [Eqs.~\eqref{EQsMotion}] into recurrence relations that incorporate both thermal fluctuations and feedback-amplified detection noise at each timestep. Applying recursive back-substitution in the long-time limit yields the following steady-state expressions for the position and momentum quadratures: 
\begin{widetext}
\begin{align}
q(n) &= \frac{2\Omega}{\bar{\gamma}_{+}-\bar{\gamma}_{-}} \lim_{n \to \infty} \sum_{j=0}^{n-1} \left(\bar{\lambda}_{-}^j - \bar{\lambda}_{+}^j\right) \left[\sqrt{2\gamma n_{\mathrm{th}}}\, dW^{\mathrm{th}}_{n-j-1} + \mathcal{A}dW^{\mathrm{det}}_{n-j-1} \right], \label{SteadyState_Position} \\
p(n) &= \frac{1}{\bar{\gamma}_{+}-\bar{\gamma}_{-}} \lim_{n \to \infty} \sum_{j=0}^{n-1} \left(\bar{\gamma}_{+} \bar{\lambda}_{+}^j - \bar{\gamma}_{-} \bar{\lambda}_{-}^j\right) \left[\sqrt{2\gamma n_{\mathrm{th}}}\, dW^{\mathrm{th}}_{n-j-1} +\mathcal{A}dW^{\mathrm{det}}_{n-j-1} \right]\;. \label{SteadyState_Momentum}
\end{align}
\end{widetext}
where the effective damping rates \(\bar{\gamma}_{\pm} = \bar{\gamma} \pm \tilde{\Omega}\) are defined in terms of \(\tilde{\Omega} \equiv \sqrt{\bar{\gamma}^{2} - 4\Omega^{2}}\), which characterizes the system’s dynamical response. Here, \(\bar{\gamma} \equiv \gamma + \beta\mathcal{C}\) represents the total effective damping rate, combining intrinsic viscous damping and phase-adaptive dissipation. The coefficients \(\bar{\lambda}_{\pm} = 1 - \Big(\frac{\bar{\gamma}_{\pm} }{2}\Big)dt\) determine the decay weights for past noise contributions, $\mathcal{A}=\eta\beta\mathcal{C}/\Omega$ scales the detection noise amplification in the feedback loop.

\subsubsection{Position power spectral density}
In our formalism, the detected oscillator position after feedback is activated is expressed as $\tilde{q}_{\mathrm{det}}(t) = q(t) + \eta W^{\mathrm{det}}$, where $q(t)$ represents the  position including thermal noise and feedback contributions and $\eta W^{\mathrm{det}}$ denotes additive detection noise. In dimensional form we can express it as
\[
z_{\mathrm{det}}(t) = \tilde{z}(t) + q_{\mathrm{zpm}} \eta\, W^{\mathrm{det}}(t),
\]

where $\tilde{z}(t)=q_{\mathrm{zpm}}q(t)$ is the modified position quadrature due to applied phase-adaptive contribution. Here $q(t)$  can be obtained by using Eq. \eqref{SteadyState_Position} in the large $n$-limit.

Within the discretized approach,  the position power spectral density (PSD) is defined by the Fourier transform:
\begin{equation}
S(\omega) = \lim_{dt \to 0} dt \sum_{n'=-\infty}^{\infty} e^{-i \omega n' dt} \langle z_{\mathrm{det}}(n) z_{\mathrm{det}}(n+n') \rangle. \label{PSD_total}
\end{equation}

Summing over $n'$ and taking the $dt \to 0$ limit yields the following form of the PSD:
\begin{widetext}
\begin{align}
S(\omega) &= \frac{2 \gamma k_B T_{\mathrm{th}} / m}{(\omega^2 - \Omega^2)^2 + (\gamma + \beta \mathcal{C})^2 \omega^2} +\left( \frac{\hbar}{m \Omega} \right)\frac{\eta^2}{\omega^2}+ \left( \frac{\hbar}{m \Omega} \right) \left[ \frac{(\eta \beta \mathcal{C})^2}{(\omega^2 - \Omega^2)^2 + (\gamma + \beta \mathcal{C})^2 \omega^2} \right]. \label{ColdDampingLimit}
\end{align}
\end{widetext}
Above position noise spectrum comprises three physically distinct components. The first contribution originates from stochastic thermal forces due to collisions with the background gas at temperature $T_{\mathrm{th}}$. This thermal noise exhibits a Lorentzian profile centered near the mechanical resonance frequency $\Omega$, with its magnitude proportional to the thermal energy $k_B T_{\mathrm{th}}$ and damping rate $\gamma$. The spectral lineshape is modified by the total effective damping $(\gamma + \beta\mathcal{C})$, representing the combined effect of intrinsic viscous drag and feedback-induced dissipation.

The second component constitutes the base detection noise floor, which dominates the off-resonant spectrum at low frequencies ($\omega \ll \Omega$). Characterized by a $1/\omega^2$ frequency dependence, this term is independent of both mechanical parameters ($\Omega$, $\gamma$) and feedback settings ($\mathcal{C}$), representing a fundamental limit determined solely by the detection noise performance $\eta$. 

The third contribution arises from detection noise amplified through the feedback loop. Scaling quadratically with both the feedback gain $\mathcal{C}$ and detection noise amplitude $\eta$, this term represents a fundamental trade-off: while higher gain enhances cooling efficiency, it simultaneously injects additional noise. This feedback-amplified component shares the Lorentzian resonance profile of the thermal noise and dominates near the mechanical resonance ($\omega \approx \Omega$), directly competing with thermal fluctuations.
\section{Improvement of detection noise}
\label{AppendixE}

\begin{figure}[ht!]
\centering
\includegraphics[width=0.6\columnwidth]{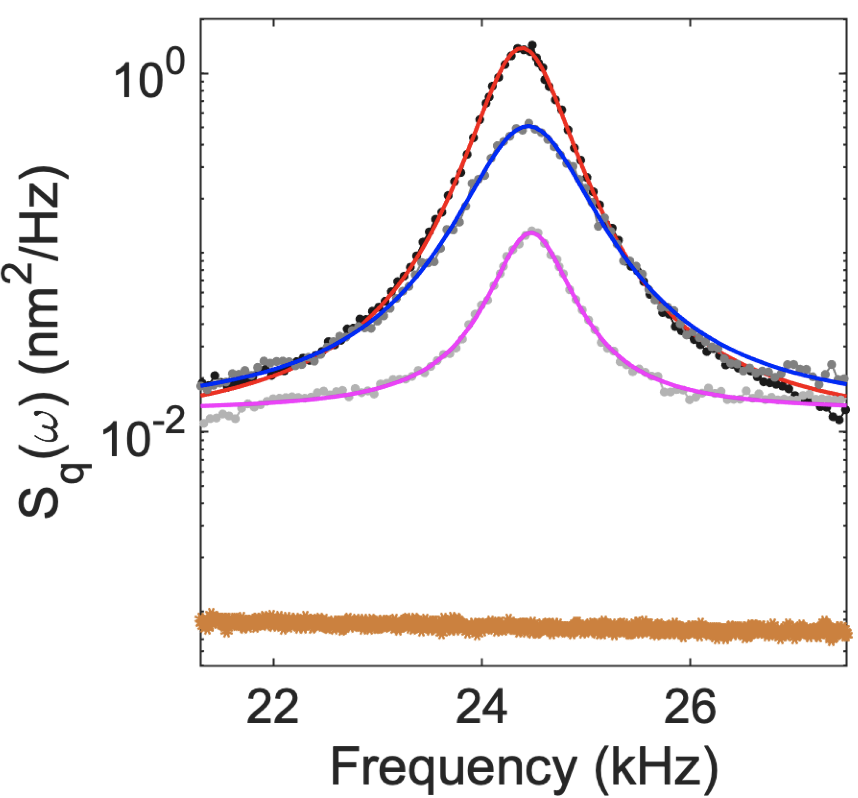}
\caption {Measured spectra of axial mechanical motion of a 400 nm silica are shown. The spectra are fitted with Eq.~\eqref{ColdDampingLimit}. The spectra without feedback cooling at 2 mbar is shown in red, with feedback cooling at 2 mbar is shown in blue and at 0.5 mbar with feedback is shown in magenta. The detection noise (26 pm/$\sqrt{\mathrm{Hz}}$)  without any particle in the optical trap and with the spatial aperture closed is shown in copper color.}
\label{400nm}
\end{figure}

A disadvantage of single-pixel detection using an amplified photodiode is that the extraneous light which comes out of the fiber cladding gets collected and amplified, setting up a strong detection noise background, ultimately contributing to the feedback noise. One way we tackled this problem is using a larger 400 nm silica particle which scatters more light due to an increase in scattering cross-section ($\sigma_{scat}\propto m^2$) and thereby allowing us to close the spatial filter in front of the photodiode, offering a lower noise floor. The detection floor in this experimental configuration is 26 pm/$\sqrt{\mathrm{Hz}}$.  The particle could be stably trapped down to 0.5 mbar. Fig.~\ref{400nm} shows the feedback cooling in action for a bigger 400 nm particle. The axial eigenfrequency is 24.7 kHz. All the spectra are fitted with Eq.~\eqref{ColdDampingLimit} for negligible $\eta$. The red fit indicates PSD of position of the oscillator at 2 mbar in the absence of feedback cooling while the blue curve indicates spectrum with feedback cooling at 2 mbar. The spectrum fitted in magenta is recorded at 0.5 mbar with feedback cooling on. The fitting parameters are $\beta=8$ $\mathrm{GHz}$, $\mathcal{C}=4.1\times10^{-7}$. The noise floor is shown in copper color.    

        %
	%
	%
	%
	%
	%


%

\end{document}